
\input harvmac
\Title{hep-th/9407106, RU-94-57}
{\vbox{\centerline{``Integrating in'' and exact superpotentials in 4d}}}
\bigskip
\centerline{Kenneth Intriligator}
\smallskip
\centerline{\it Department of Physics and Astronomy}
\centerline{\it Rutgers University, Piscataway, NJ 08855-0849}
\bigskip
\baselineskip 18pt
\noindent
We discuss integrating out matter fields and integrating in matter
fields in four dimensional supersymmetric gauge theories.
Highly
nontrivial exact superpotentials can be easily obtained by starting from
a known theory and integrating in matter.

\Date{7/94}
\noblackbox

\def\pf{{\rm Pf ~}}
\def\xt{{X_{\hat r}}}
\def\gt{{g_{\hat r}}}
\def\gG{{\cal G}}
\def\np#1#2#3{Nucl. Phys. B{#1} (#2) #3}
\def\pl#1#2#3{Phys. Lett. {#1}B (#2) #3}

\def\physrev#1#2#3{Phys. Rev. {D#1} (#2) #3}

\def\pf{{\rm Pf ~}}
\font\litfont = cmr6
\def\half{{\litfont {1 \over 2}}}
\def\gG{{\cal G}}
\lref\nonren{N. Seiberg, \pl{318}{1993}{469}.}
\lref\ils{K. Intriligator, R.G. Leigh and N. Seiberg,
RU-94-26, hep-th/9403198, Phys. Rev. D in press}
\lref\ads{I. Affleck, M. Dine, and N. Seiberg, \np{241}{1984}{493};
\np{256}{1985}{557}.}
\lref\russ{M.A. Shifman and A.I Vainshtein, \np{277}{1986}{456};
\np{359}{1991}{571}.}
\lref\cerne{G. Veneziano and S. Yankielowicz, \pl{113}{1982}{321};
T.R. Taylor, G. Veneziano, and S. Yankielowicz, \np{218}{1983}{493}.}
\lref\svgc{M.A. Shifman and A. I. Vainshtein, \np{296}{1988}{445}.}
\lref\vadim{V. Kaplunovsky and J. Louis, UTTG-94-1,
hep-th/9402005}
\lref\nati{N. Seiberg, hep-th/9402044, \physrev{49}{1994}{6857}.}
\lref\swi{N. Seiberg and E. Witten, RU-94-52, hep-th/9407087}

\newsec{Introduction}
The Wilsonian effective superpotential of a four dimensional
supersymmetric gauge
theory is highly constrained and can often be obtained exactly
\refs{\nonren , \ils}.  In
particular, symmetries constrain its functional form, dynamical
considerations dictate its singularity structure, and holomorphy then
provides the exact superpotential in all of field space by analytic
continuation.  The exact superpotentials so obtained can be highly
non-trivial, reflecting interesting non-perturbative phenomena.
By applying these techniques to a variety of examples, some conjectured
general principles have emerged.  These principles, discussed in \ils ,
have to do with the linearity
of the superpotentials in the coupling constants (see
\vadim\ for a related discussion).  We discuss how and
when it is possible to use these simple principles to easily obtain
exact superpotentials.  The technique, which we refer to as
``integrating in matter'', can be easily applied even for highly
non-trivial theories, where
the more direct considerations of \ils\ would prove very difficult.

\newsec{Integrating matter out and in}

Consider two theories.  The first, which we will refer to as the
``downstairs'' theory, is a supersymmetric gauge theory with gauge group
$\gG=\prod _s\gG_s$ and matter chiral superfields $\phi _i$ transforming
in representations $R_i$ of $\gG$.  At the classical level there are
flat directions in $\phi _i$ field space, with coordinates given by the gauge
invariant polynomials $X_r$ in the fields $\phi _i$\foot{The $X_r$ are a
finite basis of the basic gauge invariants; all gauge invariants can be
reduced to sums and products of the $X_r$.}.
The non-perturbative gauge dynamics
generates an effective superpotential $W_d(X_r, \Lambda _{s,d}^{n_{s,d}})$
for the fields $X_r$.  In this superpotential,  $\Lambda _{s,d}$ are the
scales of the $\gG_s$ gauge
dynamics in the downstairs theory; they are related to
the Wilsonian running coupling constants by the exact \russ\ 1-loop beta
functions as $\Lambda _{s,d}^{n_{s,d}}=E ^{n_{s,d}}e^{-8\pi
^2/g_s^2(E)}$.  Here
$n_{s,d}=\half(3G_s- \mu _s)$ with $G_s$ the
index of the adjoint representation of gauge group $\gG _s$
and $\mu _s\equiv \sum _i\mu ^i_s$, where $\mu ^i_s$ is the index
of the representation $R_i$ of matter field $\phi _i$ in the gauge group
$\gG _s$.

The second theory, which we refer to
as the ``upstairs'' theory, differs from the downstairs theory only in
that it contains an additional matter field $\hat \phi$ in a
representation $\hat R$ of $\gG$ and,
if $\hat R$ is not real, a conjugate field $\hat \phi ^c$ such that
the ``meson'' $\hat M=\hat \phi
\hat \phi ^c$ is gauge invariant.  The gauge invariant polynomials of
the upstairs theory are the $X_r$ of the downstairs theory along with
some additional polynomials $X_{\hat r}$.  The gauge dynamics generates
an effective superpotential $W_u(X_r, X_{\hat r}, \Lambda _s ^{n_s})$.
The scales $\Lambda _s$ of the gauge dynamics in the upstairs theory are
related to the Wilsonian running coupling constants by
$\Lambda _s^{n_s}=E^{n_s} e^{-8\pi ^2/g^2_s(E)}$, where
$n_s=n_{s,d}-\half \hat \mu _s$ with $\hat \mu _s$ the
index of $\hat R$ plus that of its conjugate
(if it is not real) in $\gG _s$.

Consider now modifying the upstairs theory by turning
 on a tree level superpotential
$W_{tree}=\sum _{\hat r}\gt\xt$ for the macroscopic
variables containing the field $\hat \phi$ or its conjugate.
We will now assume the following

$\bullet$ {\it Principle of Linearity:} \refs{\ils , \vadim} the
full superpotential $W_f$ is then simply
\eqn\wupi{W_f(X_r, \xt , \Lambda _s^{n_s}, \gt )
=W_u(X_r,\xt , \Lambda _{s}^{n_s})+\sum _{\hat
r}\gt\xt .}

\noindent
At this stage we could consider integrating out the field $\hat \phi$
and, correspondingly, the fields $\xt$.  Solving for the fields $\xt$ in
\wupi\ using their equations of motion ${\partial W_f\over \partial
\xt}=0$ will yield a
new superpotential
\eqn\wdni{W_f(X_r, \langle \xt \rangle, \Lambda _s^{n_s}, \gt )\equiv
W_l(X_r, \Lambda _s^{n_s}, g_{\hat r})=W_d(X_r,
\Lambda _{s,d}^{n_{s,d}})+W_I(X_r, \Lambda _s^{n_s}, \gt ),}
where $W_d$ is the dynamically generated superpotential of the
downstairs theory and $W_I$ is an additional superpotential, which is
irrelevant in the renormalization group sense.
In particular, $W_I\rightarrow 0$ for $\hat
m\rightarrow \infty$ where $\hat m$ is the $\gt$ for $\hat M$ in
$W_{tree}$, the mass of the field $\phi$.
Also, $W_I=0$ when $\gt =0$ for all $\gt$ besides
$\hat m$.

We now introduce a second conjectured principle

$\bullet$ {\it Simple Thresholds:} The scales $\Lambda _{s,d}$ of
the downstairs theory in \wdni\ are {\it exactly}
related to the scales $\Lambda _s$ of the upstairs theory by matching
the running couplings $g_s^2(E)$ at
the scale $E=\hat m$, independent of
all other couplings:
\eqn\udmi{\Lambda _{s,d}^{n_{s,d}}=\Lambda _s^{n_s}\hat m ^{\hat \mu
_s/2}.}
This conjecture is a generalization of the conjecture in \ils\ stating
that the superpotential with the massive fields $S_s$ integrated in is
linear in the couplings $\log
\Lambda _s^{n_s}$.  It is possible to prove these conjectures in examples
on a case by case basis but no general proof is known.
To summarize, if the
superpotential $W_u$ is known, we can ``flow'' down to the
superpotential of the downstairs theory.

Although we have integrated out the variables $\xt$ in obtaining \wdni ,
we actually have not lost any information.  This is very different from
the usual idea of coarse graining and is related to the
linearity principle.  In fact, $W_l$ is simply a Legendre transform of
$W_u$; it is possible to obtain $W_u$ from $W_l$ by an inverse Legendre
transform.  In particular, consider a theory with the superpotential
\eqn\wdnc{W_n(X_r, Y_{\hat r}, \Lambda _s^{n_s}, \gt )=
W_l(X_r, \Lambda _{s}^{n_s}, \gt )-\sum _{\hat r}\gt Y_{\hat
r},}
where $Y_{\hat r}$ are some additional gauge singlets which, as in \wupi
, can be added to
the theory without affecting $W_l$.  Now suppose that we consider the
couplings $\gt$ in the above superpotentials to be fields.  If we
integrate out the $\gt$ from \wdnc\ by solving for them using their
equations of motion ${\partial W\over\partial\gt}=0$,
we will obtain
\eqn\wupii{W_n(X_r, Y_{\hat r}, \Lambda _{s}^{n_s}, \langle \gt \rangle
)=W_u(X_r,
\xt =Y_{\hat r}, \Lambda _{s}^{n_s}).}
The equality in \wupii\ follows because we could have added the singlets
$Y_{\hat r}$ and their contribution in \wdnc\
to the original theory \wupi\ to
obtain the superpotential
\eqn\wupl{W=W_u(X_r, \xt, \Lambda _{s}^{n_s})+\sum _{\hat r}\gt(\xt
-Y_{\hat r}).}
Now the $g_r$ are simply Lagrange multipliers and, upon integrating them
out and setting $Y_{\hat r}$ to $\xt$, we have done nothing.
We thus see that the
superpotential $W_u(X_r,\xt , \Lambda _s^{n_s})$ of the
upstairs theory can be obtained from
\eqn\wnx{W_n=W_l(X_r,\Lambda _{s}^{n_{s}},\gt )-\sum _{\hat r}\gt \xt}
by integrating out the $\gt$.  The key point is that the downstairs
theory can be much simpler
than the upstairs one and hence it is often much easier to obtain $W_l$
by direct methods than it would be to directly analyze the upstairs
theory.  Nevertheless, once $W_l$ is known, we can
``flow'' up to the superpotential $W_u$ of the upstairs theory by using
\wnx .

\newsec{Theories with only quadratic gauge invariants}

When the gauge invariants $\xt$ are only quadratic,
the tree level superpotential added in \wupi\ simply
gives the field $\hat \phi$ and its conjugate a mass $\hat m$;
the superpotential $W_I=0$ in \wdni .  In theories with only quadratic
gauge invariants, it is sensible to choose pure glue $\gG$ Yang Mills as
the downstairs theory, with all matter to be integrated in using \wnx .

Consider, for example, $SU(N_c)$ QCD with $N_f<N_c$ flavors
as the ``upstairs'' theory and $SU(N_c)$ Yang-Mills as the
``downstairs'' theory.  The $X_{\hat r}$ are the mesons $M=Q\tilde Q$
in the $(N_f,N_f)$ of the global $SU(N_f)_L\times SU(N_f)_R$ flavor
symmetry.  We can decouple these fields by turning on the tree level
mass terms $W_{tree}=\Tr\ mM$, where the mass matrix $m$ is in the $(\bar
N_f, \bar N_f)$ of the flavor symmetry, to obtain pure glue $SU(N_c)$
Yang Mills theory. $SU(N_c)$ Yang Mills theory is known to have gluino
condensation\foot{In \refs{\svgc, \ils}, gaugino condensation was proven
to occur by starting with the {\it calculated} \ads\  instanton induced
superpotential for $N_f=N_c-1$ and integrating out matter.  We are here
taking the opposite route as an illustration of integrating in matter.}
which is conveniently described by the effective
superpotential for the (massive) glueball superfield $S=-W^2_{\alpha}$
\eqn\wym{W_{d}=S\left[\log \left({\Lambda _d ^{3N_c}
\over S^{N_c}}\right) +N_c\right].}
The scale $\Lambda _d$ is related to the scale $\Lambda$ of the upstairs
theory by the matching condition on the running coupling at the scales
where the fields decouple; as in \udmi , $\Lambda _d^{3N_c}=\Lambda
^{3N_c-N_f}\det m$.  Using \wnx , the superpotential of the
upstairs theory is obtained from
\eqn\wqcdi{W_n=   S\left[\log \left({\Lambda  ^{3N_c-N_f}\det
m\over S^{N_c}}\right) +N_c\right]-\Tr\ mM}
by integrating out the ``field'' $m$.  Setting
${\partial W_n\over \partial m}=0$ gives $\langle m\rangle =SM^{-1}$ and
\wqcdi\ becomes
\eqn\wintouth{W_u=   S\left[\log \left({\Lambda ^{3N_c-N_f}
\over S^{N_c-N_f}\det Q\tilde Q}\right) +N_c-N_f\right].}
Because $S$ is always massive, it should be integrated out.
Upon doing so, \wintouth\
indeed gives the correct low energy effective
superpotential of the upstairs theory \refs{\ads , \cerne}.

As another example, consider $SU(2)$ gauge theory with $2N_f$ doublets,
$Q_{ic}$ with $i=1, \dots 2N_f$ a $SU(2N_f)$ flavor index
(with $N_f<6$ for asymptotic freedom) and $c$ a $SU(2)$ color index,
as the upstairs theory.  Decoupling the doublets with the mass terms
$W_{tree}= \half m^{ij}V_{ij}$, where $m$ is the skew-symmetric mass
matrix, $V_{ij}=Q_{ic}Q_{jd}\epsilon ^{cd}$ are the gauge invariant
objects, and the $\half$ corrects for a double counting, gives
$SU(2)$ Yang-Mills as the downstairs theory.  The scales are related as
in \udmi\ by
$\Lambda ^{6-N_f}\pf m =\Lambda _d^6$.  Using \wym\ and \wnx ,
the superpotential of
the original theory is thus obtained from
\eqn\wqiia{W_n=S\left[\log \left({\Lambda ^{6-N_f}\pf m\over
S^2}\right)+2\right] -\half m^{ij}V_{ij}}
by integrating out the $m$.  Enforcing ${\partial W_n\over
\partial m}=0$ gives $\langle m\rangle=SV^{-1}$ and \wqiia\ becomes
\eqn\wqiib{W_u=S\left[\log\left({\Lambda ^{6-N_f}\over S^{2-N_f}\pf
V}\right)+2-N_f\right].}
Integrating out $S$, these $W_u$
are indeed correct, giving results explained
in \nati .

For the general case of a theory with gauge group $\gG=\prod _s\gG_s$
and matter fields $\phi _i$ such that all gauge invariants $X_r$ are
quadratic in the $\phi _i$, we can take for the downstairs
theory the
different decoupled pure glue $\gG _s$ Yang Mills theories.
The superpotential $W_l$ in \wnx\ is then simply a sum over the
decoupled gaugino condensation superpotentials, i.e. \wym\ with $N_c$
generalized to $\half G_s$, of each
$\gG _s$ Yang-Mills theory.

It is now possible to integrate in the matter fields.  To take any
non-abelian flavor symmetries into account, label the $\phi _i$ as $\phi
_{r,a}$, where $a$ is a flavor symmetry
index.  The gauge invariant objects are
the mesons $(M_r)_a^b=\phi _{r,a}\tilde \phi ^{r,b}$ and the
the mass terms are $\sum _r m_rM_r$, with the
appropriate sum over the flavor indices implicit.
The conjecture \udmi\ gives for the matching of the scales
\eqn\udmatching{\Lambda _{s,d}^{3G_s/2}=\Lambda _s^{(3G_s-\mu
_s)/2}\prod _im_i^{\mu ^i_s/2},}
where $m_i$ are the masses where the matter fields $\phi _i$ decouple.
When there are non-abelian flavor symmetries (as in the above examples)
under which the masses transform, the $m_i$ in \udmatching\ are to be
understood as the eigenvalues of the mass matrices.
The products of eigenvalues $\prod
_im_i^{\mu ^i_s/2}$ are invariant under the non-abelian symmetries and
will thus be given by products of $\det m_r$ (or $\pf m_r$ for
pseudo-real representations).  Using \wnx ,
the superpotential of the theory with matter is obtained from
\eqn\gswsm{W_n=\sum _sS_s\left[\log\left({\Lambda _s^{(3G_s-\mu
_s)/2}\prod _im_i^{\mu _s^i/2}\over S_s^{G_s/2}}\right)+\half G_s\right]
-\sum _r m_rM_r}
by integrating out the $m_r$ and the $S_s$.

As an example, consider $\prod _{s=1}^4SU(2)_s$ gauge
theory with matter content in the representations
$\phi _{1,a}=(2,1,1,1)$ for $a=1\dots 4$,
$\phi _2=(2,2,1,1)$, $\phi _3=(1,2,2,1)$ and $\phi _4=(1,1,2,2)$.
The gauge invariants are $(M_1)_{ab}=\phi
_{1,a}\phi _{1,b}$, in
the  6 of the $SU(4)$ global flavor symmetry,
$M_2=\phi _2^2$, $M_3=\phi _3^2$,
and $M_4=\phi _4^2$.  The superpotential \gswsm\ is
\eqn\extiv{\eqalign{W_n&=S_1\left[\log\left({\Lambda_1 ^3(\pf m_1)m_2\over
S_1^2}\right)+2\right]+S_2 \left[\log\left({\Lambda _2^4m_2m_3\over
S_2^2}\right)+2\right]-\half \Tr\ m_1M_1\cr
&+S_3\left[\log\left({\Lambda _3^4m_3m_4\over
S_3^2}\right)+2\right]
+S_4\left[\log\left({\Lambda _4^5m_4\over S_4^2}\right)+2\right]
 -\sum _{r=2}^4m_rM_r.\cr}}
Integrating out the $S_s$ and $m_r$ by their equations of motion gives
a superpotential $W_u=S_4-S_1$, where $S_1$ and $S_4$ are obtained by
solving
$${\Lambda_1 ^3(S_1+S_2)\over
(\pf M_1)M_2}={\Lambda _2^4(S_1+S_2)(S_2+S_3)
\over S_2^2M_2M_3}=
{\Lambda _3^4(S_2+S_3)(S_3+S_4)\over
S_3^2M_3M_4}={\Lambda _4^5(S_3+S_4)\over S_4^2M_4}=1.$$
Although this gives a superpotential which is very complicated, reflecting
some of the complicated non-perturbative
dynamics, it was obtained simply from gaugino condensation in the
decoupled
downstairs Yang-Mills theories
along with the simple matching relations \udmatching .
All of the superpotentials discussed in \ils\ can be easily obtained
using this technique.

\newsec{Theories with non-quadratic gauge invariants}

Theories with non-quadratic gauge invariants are more complicated
because the superpotentials $W_I$ in \wdni\ are nonzero.  In the
simplest case we would have $W_I=W_{tree,d}$,
the superpotential obtained from integrating $\hat
\phi$ and its conjugate out from the $W_{tree}$ in \wupi\ by their
equations of motion.  Writing $W_I=W_{tree,d}+W_{\Delta}$, it is
sometimes possible to use the
symmetries, along with the requirement that $W_{\Delta}\rightarrow 0$ in
the $\hat
m\rightarrow \infty$ limit (where only $W_d$ remains)
and also in the limit when the $\Lambda
_s\rightarrow 0$ (where only $W_{tree,d}$ remains), to argue that
$W_{\Delta}$ =0.  When this is the case, the ``integrating in'' procedure
is still useful.  Even in this case,
once a nonzero tree level superpotential
is generated, integrating out the remaining matter would result in a
terrible mess.  So we will be unable to obtain as nice of an expression
as \gswsm .   Nevertheless, the
technique of integrating in matter can be used to easily obtain
superpotentials on a case by case basis.

As an example,
consider $SU(N_c)$ QCD with $N_c$ flavors as the upstairs theory and take
$SU(N_c)$ QCD with $N_c-1$ flavors as the downstairs theory.  By a flavor
rotation we can add the mass term $mQ_N\tilde Q_N$=$mM_{NN}$ only for
the $N$-th flavor.  In addition we should add couplings for
$B=\det Q$ and $\tilde B=\det \tilde Q$ since they involve the fields
$Q_N$ and $\tilde Q_N$: $W_{tree}=mM_{NN}+bB+\tilde b \tilde B$.  The
superpotential \wnx\ is $W_n=W_d+W_{tree,d}+W_{\Delta}-W_{tree}$.
The dynamically generated superpotential of the
downstairs theory is $W_d= {\Lambda _d^{2N_c+1}\over \det M_d}$,
where $M_d$ are the mesons involving the $N_c-1$ flavors of the
downstairs theory.  Integrating
out the fields $Q_N$ and $\tilde Q_N$
{}from $W_{tree}$ gives $W_{tree,d}= -{b\tilde b\over
m}\det M_d$.
The symmetries determine \nonren\
$W_{\Delta}={b\tilde b\over m}\det M_df({
\Lambda _d^{2N_c+1}m\over b\tilde b\det M_d^2})$
where, because the gauge group is completely broken by $\det M_d$,
$f(u)=\sum _{n=0}^\infty
a_nu^n$.  Adjusting the relative strength of the limits
$m\rightarrow
\infty$ and $\Lambda _d \rightarrow 0$, where it is known that
$W_{\Delta}\rightarrow 0$, shows that $W_{\Delta}=0$ everywhere.
The matching condition \udmi\ on the scales is
$\Lambda _d^{2N_c+1}=m\Lambda ^{2N}$, independent of $b$ and $\tilde b$.
It is possible to prove this: again,
the symmetries would allow the equality to be modified
by a function
$g\left({\Lambda ^{2N_c+1}m\over b\tilde b \det M_d ^2}\right)$.
We know $g\rightarrow 1$ for $m\rightarrow \infty$ and also for
$\det  M_d\rightarrow \infty$, where the theory is very weakly coupled.
Adjusting the relative
strength of these limits gives $g$=1 identically.

To summarize, we have obtained for the superpotential \wnx\
\eqn\wnnmb{W_n={m\Lambda ^{2N_c}\over \det  M_d}-{b\tilde b\over m}\det
M_d-mM_{NN}-bB-\tilde b\tilde B.}
The superpotential of the upstairs theory is obtained from \wnnmb\ by
integrating out $m$, $b$, and $\tilde b$.  Doing so yields
\eqn\wnn{W_u=0\qquad\hbox{with}\qquad \det M-B\tilde B=\Lambda ^{2N},}
where we substituted the flavor invariant quantity $\det M$ for the
quantity $M_{NN}\det M_d$, obtained because of our particular choice of
integrating out the $N$-th flavor.  This
is indeed the quantum deformed
moduli space of vacua obtained (by similar reasoning) in \nati .

As another example, consider $SU(2)_L\times SU(2)_R$ gauge theory
with matter in the representations
$Q=(2,2)$, $L_{\pm}=(2,1)$ and $R_{\pm}=(1,2)$.
Without the field $Q$, this would be the two decoupled $SU(2)_L$ and
$SU(2)_R$ gauge theories, each with a single flavor; we will take this
as the downstairs
theory: $W_d=
{\Lambda _{L,d}^5\over X_L}+{\Lambda _{R,d}^5\over
X_R}$, where $X_L=L_+L_-$ and $X_R=R_+R_-$.
To get from the upstairs theory to
this downstairs theory we would add
the tree level superpotential $W_{tree}=m_QX_Q+\vec \lambda \cdot \vec Z$
where $X_Q=Q^2$ and $\vec Z=LQR$, in the $(2,2)$ representation
of the global $SU(2)\times
SU(2)$ flavor symmetry.  Integrating out the field $Q$ at tree level
gives $W_{tree,d}=-{\vec \lambda ^2\over 4m_Q}X_LX_R$.
The superpotential \wnx\ is given by $W_n=W_d+W_{tree,d}+W_{\Delta}-
W_{tree}$.
The symmetries can be used
\refs{\nonren , \ils} to show $W_{\Delta}={\vec \lambda ^2\over
m_Q}X_LX_Rf({m_Q\Lambda _{L,d}^5\over
\vec \lambda ^2 X_L^2X_R},{\Lambda _{R,d}^5X_L\over \Lambda _{L,d}^5
X_R})$.
Because the gauge group is completely broken for nonzero $X_L$ and
$X_R$, $f(u,v)=\sum _{n=0}^{\infty}\sum _{m\leq
n}a_{n,m}u^nv^m$.  Further,  $W_{\Delta}\rightarrow 0$ in the limits
$m_Q\rightarrow \infty$ or $\Lambda _L$, $\Lambda
_R\rightarrow 0$.  Adjusting the relative strength of these limits gives
$W_{\Delta}=0$.
The matching \udmi\ of the scales is $\Lambda
_{L,d}^5 =\Lambda _L^4m_Q$ and $\Lambda _{R,d}^5=\Lambda _R^4m_Q$,
independent of $\vec \lambda$.  Again, it is here possible to prove this
using the symmetries and the behavior in different limits.
Using \wnx , the effective superpotential of the upstairs theory is
thus obtained from
\eqn\wnqlr{W_n= {\Lambda _L^4m_Q\over X_L}+{\Lambda _R^4m_Q\over
X_R}-{\vec \lambda ^2\over 4m_Q}X_LX_R-m_QX_Q-\vec\lambda \cdot \vec Z,}
by integrating out $m_Q$ and $\vec \lambda$.  Doing so yields
\eqn\wuqlr{W_u=0\qquad\hbox{with}\qquad X_QX_LX_R-\vec Z^2=X_L\Lambda
_R^4+X_R\Lambda _L^4.}

As another derivation of \wuqlr , take the theory without the
fields $L_{\pm}$ as the downstairs one.  The dynamically
generated superpotential
of this downstairs theory can be determined using \gswsm\
and was discussed in detail in \ils :
$W_d=\Lambda _{L,d}^5X_R/(X_QX_R-\Lambda
_{R,d}^4)$.  In addition, there is a constraint in the downstairs theory
$X_QX_R-(QR_+)(QR_-)=\Lambda _{R,d}^4$.  This downstairs theory is obtained
{}from our upstairs one by adding
$W_{tree}=m_LX_L+\vec \lambda \cdot \vec Z$ and integrating out the
fields $L_{\pm}$.  Integrating out $L_{\pm}$ at tree level gives
$W_{tree,d}= -{\vec \lambda
^2\over 4m_L}(QR_+)(QR_-)=-{\vec \lambda
^2\over 4m_L}(X_QX_R-\Lambda _{d,R}^4)$,
where we used the mentioned constraint.
Again, the symmetries and the limiting behaviors determine
$W_{\Delta}=0$.
The scales match as $\Lambda _{L,d}^5=m_L\Lambda _L^4$ and $\Lambda
_{R,d}^4=\Lambda _R^4$.  Using \wnx , $W_u$ can be obtained from
\eqn\wnqr{W_n={m_L\Lambda _{L}^4X_R\over X_QX_R-\Lambda _R^4}-{\vec \lambda
^2\over 4m_L}(X_QX_R-\Lambda _R^4)-m_LX_L-\vec \lambda \cdot \vec Z}
by integrating out $m_L$ and $\vec
\lambda$.  Doing so indeed reproduces the same
result \wuqlr .

Although in these examples we could argue that $W_{\Delta}=0$, this is
not always the case.  In fact, whenever
$\mu _s >G_s$ for some $\gG _s$, the symmetries and limiting behaviors
can allow a nonzero
$W_{\Delta}$.  A more direct and involved analysis
would then be necessary.

\newsec{Conclusions and limitations}

We have discussed how matter can
be integrated in, as well as integrated out.  For theories with only
quadratic gauge invariants, this allowed us to obtain the general
expression \gswsm .  For theories with non-quadratic gauge invariants, the
technique of integrating in matter is useful when it is possible to
determine the obstructing superpotential $W_{\Delta}$.

An obvious general limitation of the technique of integrating in matter
is that only non-chiral matter can be so integrated in.
This is unfortunate since chiral theories are quite important: they can
dynamically break supersymmetry.  For these theories, at present, it is
necessary to conduct the detailed analysis discussed in \ils\ on a
case by case basis.

We should mention an important place where integrating
in seemingly breaks down.  Consider, for example,
$SU(2)$ gauge theory with matter $\phi$ in
the adjoint as the upstairs theory.  Adding a mass
term $m\phi ^2$ to the upstairs theory and integrating out $\phi$ gives
$SU(2)$ Yang-Mills as the downstairs theory, with scale matching $\Lambda
_d^6=\Lambda ^4m^2$.  Applying our integrating in procedure then
suggests that the upstairs theory is
described by the superpotential $W_u=0$ with a
constraint $\phi ^2=\pm 2 \Lambda ^2$ -- this is incorrect: $\phi ^2$ is
only so constrained for $m\neq 0$ \swi .  This appears to
contradict principle \wupi\ but that is only because there is more to
the story.  As discovered in \swi ,
there are additional matter fields, magnetic monopoles, which must be
taken into account in order to properly describe this upstairs theory
when $m$=0.  The moral is that it is important to take care to correctly
identify the spectrum of light fields
involved in the low energy effective theory.

Finally, our analysis was based on assuming the principles \wupi\ and
\udmi .  It would be useful to better understand their veracity.

\centerline{{\bf Acknowledgements}}

I would like to thank R. Leigh and N. Seiberg for stimulating discussions
and helpful comments.
This work was supported in part by DOE grant \#DE-FG05-90ER40559.

\listrefs

\end